\documentclass[twocolumn,showpacs,nofootinbib]{revtex4-1}

\usepackage{amsmath}
\usepackage{graphicx}

\def\to{\rightarrow}

 \def\CQG{{\it Class. Quant. Grav.} }
  \def\GRG{{\it Gen. Rel. Grav.} }  \def\IJMP{{\it Int. J. Mod. Phys.} }

  \def\MNRAS{{\it Mon. Not. R. Ast. Soc.} } 
 
 \def\PL{{\it Phys. Lett.} } \def\PR{{\it
		Phys. Rev.} } \def\PRL{{\it Phys. Rev. Lett.} } \def\PRTS{{\it
		Phys. Rept.} }

\def\al{\alpha}

\def\La{\Lambda}   
   \def\cl{{\cal L}}

 \def\frac#1#2{{\textstyle{{#1}\over
			{#2}}}} 
\def\lsim{\mathrel{\rlap{\lower4pt\hbox{\hskip1pt$\sim$}}
		\raise1pt\hbox{$<$}}}
\def\gsim{\mathrel{\rlap{\lower4pt\hbox{\hskip1pt$\sim$}}
		\raise1pt\hbox{$>$}}} \def\sqr#1#2{{\vcenter{\vbox{\hrule
				height.#2pt \hbox{\vrule width.#2pt height#1pt \kern#1pt \vrule
					width.#2pt} \hrule height.#2pt}}}}
\def\square{\mathchoice\sqr66\sqr66\sqr{2.1}3\sqr{1.5}3}
\def\beq{\begin{equation}} \def\eeq{\end{equation}}
\def\beqa{\begin{eqnarray}} \def\eeqa{\end{eqnarray}}
\def\ca{\mathcal{A}} \def\cb{\mathcal{B}} \def\cc{\mathcal{C}} \def\cd{\mathcal{D}}    \def\cl{\mathcal{L}} 

\usepackage{hyperref}

\begin{document}

\title{Dynamical analysis of generalised $f(R,\cl)$ theories}

\author{R.P.L. Azevedo}
\email[]{up201109025@fc.up.pt}
\author{J. P\'aramos}
\email[]{jorge.paramos@fc.up.pt}
\affiliation{Departamento de F\'\i sica e Astronomia and Centro de F\'isica do Porto, \\Faculdade de Ci\^encias da Universidade do Porto,\\Rua do Campo Alegre 687, 4169-007 Porto, Portugal}

\date{\today}

\begin{abstract}
In this work, we use a dynamical system approach to analyse the viability of $f(R,\cl)$ candidates for dark energy. We compare these with nonminimal coupled $f(R)$ theories and study the solutions for exponential and power-law forms in order to constraint the allowed range of model parameters.

\end{abstract}


\maketitle

\section{Introduction}\label{sec:intro}

General Relativity (GR) is currently the most well supported theory of gravitation, boasting an enormous body of experimental evidence \cite{experimental}, from the accurate prediction of Mercury's orbit to the more recent detection of gravitational waves \cite{ligo}. In spite of this, recent observations \cite{dark} are incompatible with a purely baryonic matter content and must resort to exotic non-baryonic forms of matter and dark energy to accurately model the rotation of galaxies and the accelerated expansion of the Universe.

It is in this context that other models began to appear and attempt to explain this large scale behaviour. Among the most prominent phenomenological proposals are the so-called $f(R)$ theories \cite{felice,fR}, where the Einstein-Hilbert action is replaced by a nonlinear function of the scalar curvature, and models that present nonminimal couplings (NMC) between matter and curvature \cite{early,Bertolami:2007gv}. Some of these models have been shown to be able to mimic dark matter \cite{NMCDM}, dark energy \cite{Bertolami:2010cw,Bertolami:2011fz,Bertolami:2013uwl} and explain post-inflationary preheating \cite{Bertolami:2010ke} and cosmological structure formation \cite{NMCS}.

Previous attempts at solving these cosmological problems using a NMC model have resorted to a coupling between curvature and a scalar field \cite{singleNMCDE,repetido,singleNMCCI,singleNMCHI}, but did not extend this coupling to the baryonic matter content. More recently, a dynamical system analysis approach was used to analyse a model that incorporated both $f(R)$ theories and a NMC with the baryonic matter content \cite{rafael2014}.

Following the renewed interest on $f(R)$ theories over the past decade and the previous work reported in Ref. \cite{rafael2014}, in this work we use a dynamical system approach to study a more general $f(R,\cl)$ group of theories \cite{lobo2010}, that allow for more nonlinear couplings between matter and curvature. We use this method to check the viability of several models on large scales, such as possible candidates for dark energy. Other similar studies, albeit in a different context, can be found in Refs. \cite{Azizi:2014qsa,Carloni:2004kp,Carloni:2007br,fttg}.

This work is organized as follow: the general NMC $f(R,\cl)$ model is discussed in Sec. \ref{sec:model}; the derivation of the corresponding dynamical system is found in Sec. \ref{sec:system}; confirmation of the results obtained in GR and previous work are presented in Secs. \ref{sec:GR} and \ref{sec:f1+f2}, respectively; the results and respective discussion of an exponential and a power law models can be found in Secs. \ref{sec:exp} and \ref{sec:powerlaw}, respectively. Finally, the conclusions are presented in Sec. \ref{sec:conclusion}. Information on the relevant physical quantities can be found on Appendix \ref{app:physicalq}, while an analysis of the necessary conditions for a de Sitter Universe can be found in Appendix \ref{app:desitter}.

\section{The Model}\label{sec:model}

We consider a broad generalization of the Einstein-Hilbert action that follows from $f(R)$ theories \cite{felice}, but allows for an arbitrary non-minimal coupling between matter and curvature, embodied in the action,
\begin{equation}\label{action}
S = \int d^4x \sqrt{-g} ~f\left(R,\cl\right),
\end{equation}
\noindent where $f(R,\cl)$ is an arbitrary function of the scalar curvature $R$ and the matter Lagrangian density $\cl$ and $g$ is the determinant of the metric.

We may recover GR by setting $f(R,\cl) = \kappa (R-2\Lambda)+\cl$ with $\kappa=c^4/(16\pi G)$; similarly, $f(R)$ theories are given by assuming the separation $f(R,\cl) = f(R) + \cl$, while NMC theories posit the form $f(R,\cl) = f_1(R) + f_2(R) \cl$.

The field equations are obtained by a null variation of the action with respect to the metric, and take the form
\begin{equation}\label{field}
f^R G_{\mu\nu} = {1\over 2}g_{\mu\nu}\left(f-f^R R\right) + \Delta_{\mu\nu}f^R + {1 \over 2} f^L \left(T_{\mu\nu}-g_{\mu\nu}\cl \right),
\end{equation}
\noindent where
\begin{equation}\label{fderiv}
f^{\overbrace{R\dots R}^n \overbrace{L\dots L}^m} \equiv {\partial^{n+m} f(R,\cl)\over \partial R^n \partial \cl^m},
\end{equation}
\noindent $\Delta_{\mu\nu} \equiv \nabla_\mu \nabla_\nu - g_{\mu\nu} \square$, and the matter energy-momentum tensor is defined as
\begin{equation}\label{energy-mom}
T_{\mu\nu} = - {2 \over \sqrt{-g}}{\delta\left(\sqrt{-g}\cl\right) \over \delta g^{\mu\nu}}.
\end{equation}
We can take the covariant derivative of the field equations and use the first Bianchi identity to obtain the conversation law for the energy-momentum tensor,
\begin{equation}\label{conserv}
\nabla^\mu T_{\mu\nu} = \left(g_{\mu\nu}\cl-T_{\mu\nu}\right)\left({f^{RL}\over f^L}\nabla^\mu R+ {f^{LL} \over f^L}\nabla^\mu\cl\right),
\end{equation}
\noindent showing that the latter is no longer covariantly conserved. The above expands upon the result obtained for NMC theories \cite{Bertolami:2007gv}, which read
\begin{equation}\label{conservNMC}
\nabla^\mu T_{\mu\nu} = {f_2^{R}\over f_2}\left(g_{\mu\nu}\cl-T_{\mu\nu}\right)\nabla^\mu R.
\end{equation}
In order to analyse the present day evolution of the Universe, we consider a flat Universe and adopt the Friedman-Robertson-Walker (FRW) metric, given by
\begin{equation}\label{line}
ds^2=-dt^2+a^2(t)dV^2,
\end{equation}
\noindent where $a(t)$ is the scale factor and $dV$ is the volume element. Matter is assumed to behave as a perfect fluid, with an energy-momentum tensor
\begin{equation}\label{fluid}
T^{\mu\nu}=(\rho+P)u^\mu u^\nu + p g^{\mu\nu},
\end{equation}	
\noindent derived from the Lagrangian density $\cl=-\rho$ (see Ref. \cite{fluid} for a discussion), where $\rho$ and $p$ are the respectively the energy density and pressure of the perfect fluid, and $u^\mu$ is its four-velocity.

By substitution, this Lagrangian density and energy-momentum tensor in the conservation equation \eqref{conserv}, we once again obtain the usual continuity equation
\begin{equation}\label{contin}
\dot{\rho}+3H(1+w)\rho=0,
\end{equation}	
\noindent where $H=\dot{a}/a$ is the Hubble parameter and $w=p/\rho$ is the equation of state (EOS) parameter. Notice that, although Eq. (\ref{conserv}) implies that, in general, energy is not conserved, it turns out that assuming a FRW metric makes the factor $g_{\mu\nu}\cl-T_{\mu\nu}$ vanish: this was also the case in NMC models, as can be seen from Eq. (\ref{conservNMC}).

Substituting the metric into the field equations, we obtain the modified Friedmann equation from the $00$ component
\begin{align}\label{fried}
H^2 = {1\over 3f^R}\Big[ & {1\over 2}f^R R-3Hf^{RR} \dot{R}-{1\over 2}f - \nonumber \\
& 9H^2 f^{RL}(1+w)\rho\Big],
\end{align}
\noindent and the modified Raychaudhuri equation
\begin{align}\label{ray}
&& 2\dot{H}+3H^2 = \nonumber \\ && {1\over 2f^R}\Big[f^R R -f &-f^L(1+w)\rho -2\ddot{f^R}-4H\dot{f^R}\Big].
\end{align}
\section{Dynamical System}\label{sec:system}

We can explore solutions to the field equations by analysing the dynamical system composed of the dimensionless variables
\begin{align}\label{variables}
&x=-{f^{RR} \dot{R}\over f^R H}~~~~,~~~~y={R \over 6H^2}~~~~,~~~~z=-{f\over 6f^R H^2}, \nonumber \\
&\phi=-{3(1+w)f^{RL}\rho \over f^R}~~~~,~~~~\theta={(1+w)f^L\rho \over 2f^R H^2},
\end{align}
so that the modified Friedmann equation \eqref{fried} becomes
\begin{equation}\label{mfe}
1=x+y+z+\phi,
\end{equation}
\noindent acting as an algebraic restriction to the phase space.

It is useful to calculate the following quantities in terms of the variables \eqref{variables}:
\begin{align}\label{fdot}
&{\dot{f^R} \over f^R H} =-(x+\phi),\\ \nonumber 
&{\ddot{f^R} \over f^R H^2} = (x+\phi)(x+\phi+2-y)-{dx \over dN}- {d\phi \over dN},
\end{align}
\noindent where $N=\ln a$ is the number of e-folds, so that
\begin{equation}
{d \over dN} ={1\over H} {d \over dt}.
\end{equation}

By substituting the equalities \eqref{fdot} in Eq. \eqref{ray}, one obtains the Raychaudhuri Eq. in the dimensionless form
\begin{equation}\label{dimensionlessRaychaudhuri}
{dx \over dN} + {d\phi \over dN} = (x+\phi)(x+\phi-y)-y-3z+\theta-1.
\end{equation}
Going forward, it is useful to define the following dimensionless parameters:
\begin{align}\label{parameters}
\eta_R&={f^R R \over f},	&	\gamma_R&={f^{RL} R \over f^L},	&	\gamma_L&=-{f^{LL} \rho  \over f^L}, \nonumber \\
\alpha_R&={f^{RR} R \over f^R},	&	\beta_R&={f^{RRL} R \over f^{RL}},	&	\beta_L&=-{f^{RLL} \rho \over f^{RL}}.
\end{align}

The dynamical system consists of the evolution equations for the variables \eqref{variables} (with respect to $N$), which are now derived. Using the continuity equation \eqref{contin}, we obtain
\begin{equation}\label{evophi}
{d\phi\over dN} = \phi\left[x\left(1-{\beta_R \over \alpha_R}\right)- 3(1+w)\left(\beta_L+1\right)+\phi\right].
\end{equation}

Using \eqref{evophi}, the modified Raychaudhuri equation yields
\begin{align}\label{evox}
{dx \over dN} =& x\left[x-y+\phi\left(1+{\beta_R \over \alpha_R}\right)\right] -1 -y -3z +\theta + \nonumber \\
& \phi\left[3(1+w)\left(\beta_L\right)-y\right].
\end{align}
%ea
Differentiating the remaining variable with respect to the number of e-folds $N$, we obtain the following dynamical system
\begin{equation}\label{system1}
\begin{cases}{dx \over dN} = &x\left[x-y+\phi\left(1+{\beta_R \over \alpha_R}\right)\right] -1 -y -3z + \theta \\ &+ \phi\left[3(1+w)\left(\beta_L+1\right)-y\right] \\
{dy \over dN} = & y\left[2(2-y)-{x \over \alpha_R}\right] \\
{dz \over dN} = & z\left[x+\phi+2(2-y)\right] +{xy \over \alpha_R} -\theta \\
{d\phi \over dN} = & \phi\left[x\left(1-{\beta_R \over \alpha_R}\right)- 3(1+w)\left(\beta_L+1\right)+\phi\right] \\
{d\theta \over dN} = & {xy\phi \over \alpha_R} +\theta\left[x-2y+\phi+1-3w-3(1+w)\gamma_L\right]
\end{cases}
\end{equation}
\noindent subject to the constraint \eqref{mfe}. Applying this constraint we can eliminate one of the equations from the system: we choose to eliminate $z$ and are left with
\begin{equation}\label{system2}
\begin{cases}
	{dx \over dN} = & x\left[x-y+\phi\left(1+{\beta_R \over \alpha_R}\right)+3\right] -4 +2y +\theta \\ & + \phi\left[3(1+w)\left(\beta_L+1\right)+3-y\right] \\
	{dy \over dN} = & y\left[2(2-y)-{x \over \alpha_R}\right] \\
	{d\phi \over dN} = & \phi\left[x\left(1-{\beta_R \over \alpha_R}\right)- 3(1+w)\left(\beta_L+1\right)+\phi\right] \\
	{d\theta \over dN} = & {xy\phi \over \alpha_R} +\theta\left[x-2y+\phi+1-3w-3(1+w)\gamma_L\right]
\end{cases}.	
\end{equation}

Solving this system usually also requires writing the scalar curvature $R$ and energy density $\rho$ as functions of the variables \eqref{variables}. Failure to do so may severely limit the ensuing analysis even if the fixed points of the system (\ref{system2}) are determined, as one may not be able to translate between the assumed dimensionless variables and the physically significant quantities. A brief note on obtaining these relations can be found in Appendix \ref{app:physicalq}.
\section{General Relativity}\label{sec:GR}
We now consider the case of GR, where $f(R,\cl)=\kappa (R-2\Lambda)+\cl$, so that
\begin{align}\label{derivGR}
f^R&=\kappa, & f^L&= 1, \nonumber \\
\beta_L&=0, &\gamma_L&=0,  \nonumber \\
{x\over\alpha_R}&=-{\dot{R}\over HR}, & x\phi{\beta_R \over\alpha_R}&=0,
\end{align}
\noindent and higher order derivatives of $f(R,\cl)$ are null, thus implying $x= \phi=0$ and $\theta=4-2y=3(1+w)$. Note that even though some of the parameters in Eq. \eqref{parameters} might diverge, the combined terms appearing in the dynamical system (\ref{system2}) do not. Furthermore, the Raychaudhuri Eq. (\ref{dimensionlessRaychaudhuri}) implies that
\begin{equation} 0 = 1 + y + 3z -\theta \to y = 2 - {\theta \over 2},\end{equation}
using the algebraic constraint $z = 1 - y$ resulting from Eq. (\ref{mfe}). This allows us to read the value for $y$ and $z$, with the dimensionless variable $\theta$ determined from the differential equation arising from dynamical system (\ref{system2}),
\begin{equation}\label{systemGR}
	{d\theta \over dN} = \theta [\theta -3(1+w)],
\end{equation}
which has the solution
\begin{equation}\theta(a)={3(1+w) \over 1 + {2\kappa \Lambda \over \rho_0} \left({a\over a_0}\right)^{3(1+w)}},\end{equation}
where $\rho_0 $ is the density when $a = a_0$.

Thus, for a sufficiently small scale factor we obtain a matter dominated Universe, corresponding to the unstable fixed point $\ca$, depicted on Table \ref{fixGR}.

Conversely, for late times the stable fixed point $\cb$ is attained, so that $ q = -1$, {\it i.e.} a asymptotically De Sitter Universe with infinitely diluted matter, as expected. From definition (\ref{variables}) arises the usual relation between the expansion rate and the cosmological constant,
\begin{equation}H_0^2 = {\La \over 3 (y+z)} = {\La \over 3}. \end{equation}

\begin{table}
\caption{Fixed point and respective solutions for General Relativity.\label{fixGR}}
\begin{tabular*}{\columnwidth}{@{\extracolsep{\fill}}lcccc}
\hline \hline
&$(x,y,z,\phi,\theta)$	&$a(t)$	&$\rho(t)$	&$q$	\\ \hline 
$\mathcal{A}$	&$\left(0,{{1-3w}\over{2}},{{1+3w}\over{2}},0,3(1+w)\right)$	&$\left({{t}\over {t_0}}\right)^{{2}\over {3(1+w)}}$ 	&$\rho_0 \left({{t}\over {t_0}}\right)^{-2}$	&${{1+3w}\over{2}}$ \\
$\mathcal{B}$	&$(0,2,-1,0,0)$	&$e^{H_0 t}$ 	&$0$	&$-1$ \\
\hline \hline
\end{tabular*}
\end{table}

\section{Nonminimally coupled theories}\label{sec:f1+f2}

Having used the standard scenario posited by GR as a sanity check for our method, we now proceed and analyse the case of a combination of $f(R)$ theories and a non-minimal coupling, defined by the function $f(R,\cl)=\kappa f_1(R)+f_2(R)\cl$, in order to confirm the results obtained in Ref. \cite{rafael2014}. In the latter, the relations \eqref{Rrho} yield the additional constraint
\begin{equation}
y\left[{\phi \over 3(1+w)}-1\right]=\left[z-{\theta \over 3(1+w)}\right]{f^R_1 R\over f_1},
\end{equation}
\noindent or equivalently,
\begin{equation}
\hat{y}\left[{\hat{\Omega}_2 \over 3(1+w)}-1\right]=\hat{z}\hat{\alpha}_1,
\end{equation}
\noindent where $\hat{\alpha}_1=f^R_1 R/f_1$ and the variables $(\hat{x},\hat{y},\hat{z},\Omega_1,\Omega_2)$ are the ones used in Ref. \cite{rafael2014}; these are related to those defined in \eqref{variables} by
\begin{eqnarray}
\hat{x}=x~~~~,~~~~\hat{y}=y~~~~&,&~~~~\hat{z}=z-{\theta \over 3(1+w)}, \nonumber \\
\hat{\Omega}_1={\theta \over 3(1+w)}~~~~&,&~~~~\hat{\Omega}_2=\phi.
\end{eqnarray}
This is the same condition obtained from the modified Raychaudhuri Eq. in Ref. \cite{rafael2014}. Replacing these variables in the equation system \eqref{system1}, we get
	\begin{equation}\label{system_old}
	\begin{cases}
	{d\hat{x} \over dN} =& \hat{x}\left[\hat{x}-\hat{y}+\hat{\Omega}_2\left(1+{\hat{\alpha}_2 \over \hat{\alpha}}\right)\right] -1 -\hat{y} -3\hat{z} \\ & + 3w\hat{\Omega}_1 +\hat{\Omega}_2\left[3(1+w)-\hat{y}\right] \\
	{d\hat{y} \over dN} =& \hat{y}\left[2(2-\hat{y})-{\hat{x} \over \hat{\alpha}}\right] \\
	{d\hat{z} \over dN} =& \hat{z}\left[\hat{x}\left(1-{\hat{\alpha}_1 \over \hat{\alpha}} \right)+\hat{\Omega}_2+2(2-\hat{y})\right] \\
	{d\hat{\Omega}_1 \over dN} =& {\hat{\Omega}_2 \hat{x}\hat{y} \over 3\hat{\alpha}(1+w)} +\hat{\Omega}_1\left(1-3w+\hat{x}+\hat{\Omega}_2-2\hat{y}\right) \\
	{d\hat{\Omega}_2 \over dN} =& \hat{\Omega}_2\left[x\left(1-{\hat{\alpha}_2 \over \hat{\alpha}}\right)- 3(1+w)+\hat{\Omega}_2\right]
	\end{cases},
	\end{equation}
with $\hat{\alpha}=f^{RR} R/f^R$ and $\hat{\alpha}_2=f^{RR}_2 R/f^R_2$ --- which is exactly the system obtained in the aforementioned study, as expected.

\section{Exponential $f(R,\cl)$}\label{sec:exp}

Having established the soundness of the method here reported by comparison with GR and NMC theories, we now proceed to study the more complex model
\begin{equation}\label{expo}
f(R,\cl)=M^4\exp\left({R \over 6H_0^2}+{\cl \over 6H_0^2\kappa }\right).
\end{equation}
\noindent where $M$ is a type mass scale and $H_0^2$ will turn out to be the expansion rate of the fixed point associated with a de Sitter solution. This is the same model given as an example of $f(R,\cl)$ theories in Ref. \cite{lobo2010}. Notice that this does not simplify to General Relativity with a cosmological constant for small $R $ and $\cl$, as the constant term corresponding to the cosmological constant would appear with the wrong sign: however, the fixed points obtained below do not obey the latter conditions.

The quantities defined in Eq. (\ref{parameters}) now read
\begin{eqnarray}\label{parametersExp}
\alpha_R = \beta_R = \gamma_R = \eta_R&=& { R \over 6H_0^2} = -{y \over z}, \\ \beta_L = \gamma_L&=& -{ \rho  \over 6 H_0^2\kappa} = {\phi \over 3(1+w)} , \nonumber
\end{eqnarray}
while the constraints (\ref{Rrho}) yield
\begin{equation}\label{RrhoExp}
R = -6H_0^2 {y \over z} ~~~~,~~~~\rho = -{2 H_0^2 \kappa\over 1+w} {\theta \over z}~~~~,~~~~ \theta = \phi z.
\end{equation}
The fixed points of the dynamical system (\ref{system2}) are shown in Table \ref{fix_exp}, along with the corresponding solutions.
\begin{table}
	\caption{Fixed points and respective solutions for an exponential $f(R,\cl)$ function.\label{fix_exp}}
	\begin{tabular*}{\columnwidth}{@{\extracolsep{\fill}}lcccc}
		\hline \hline
		Point&$(x,y,z,\phi,\theta)$	&$a(t)$	&$\rho(t)$	&$q$	\\ \hline 
		$\mathcal{A}$	&$\left(1,0,0,0,0 \right)$	&$\left({t \over t_0}\right)^{1 \over 2}$ 	&$\rho_0 \left({t \over t_0}\right)^{-{3 \over 2}(1+w)}$	&$-1$ \\
		$\mathcal{B}$	&$\left(0,2,-1,0,0\right)$	&$e^{H_0 t}$ 	&$\rho_0 e^{-3(1+w)H_0 t}$	&$-1$ \\
		\hline \hline
	\end{tabular*}
\end{table}
\subsection{Point $\mathcal{A}$}\label{subsec:expA}
This solution is a saddle point that is only valid at late times, $t\rightarrow\infty$: since $y=0$, Eq. \eqref{Ricci} implies that the scalar curvature vanishes (or equivalently, the relation $R \sim y/z $ found above); however, $x\neq 0$: from its definition (\ref{variables}), this is only possible if either $f^R=0$ or $H=0$. The former implies
\begin{align}
f^R&={M^4\over6H_0^2}\exp\left({R \over 6H_0^2}-{\rho \over 6H_0^2\kappa}\right) \nonumber \\
&= {M^4\over6H_0^2}\exp\left(-{\rho \over 6H_0^2\kappa}\right)= 0,
\end{align}
which leads to the unphysical result $\rho \rightarrow \infty$.

The vanishing scalar curvature implies that $H(t) = 1/2t$: the latter condition $H=0$ thus implies that $ t\rightarrow\infty$, as mentioned previously.

\subsection{Point $\mathcal{B}$}\label{subsec:expB}

This is a saddle point where $\rho\rightarrow 0$ as $t\rightarrow\infty$, {\it i.e.} an empty Universe. This results in a de Sitter solution with an exponential scale factor but, unlike the case of GR, it is unstable due to the exponential form of $f(R,\cl)$. Interestingly, it could imply that the  current phase of accelerated expansion is temporary.

The definition (\ref{variables}) now reads
\begin{equation}
z=-{H_0 \over H} = -1,
\end{equation}
so that the expansion rate is indeed given by $H=H_0$.

\section{Power law $f(R,\cl)$}\label{sec:powerlaw}

We now consider the model
\begin{equation}
f(R,\cl)=\left(\kappa M^2\right)^{-\varepsilon}\left(\kappa R+\cl\right)^{(1+\varepsilon)},
\end{equation}
\noindent where $M$ is a characteristic mass scale and $\varepsilon\ll 1$, so that it represents a power law extension of GR.

In this model, the quantities defined in Eq. (\ref{parameters}) become
\begin{eqnarray}\label{parametersPower}
{\alpha_R \over \varepsilon} = {\gamma_R \over \varepsilon} =  {\eta_R \over \varepsilon + 1} = {\beta_R \over \varepsilon - 1} &=&  {3 (1+w)y \phi \over 3(1+w)y \phi - \theta } , \nonumber \\ \beta_L = \left( 1 - { 1 \over \varepsilon}\right) \gamma_L &=&  {(1-\varepsilon)\theta \over 3(1+w)y\phi - \theta},
\end{eqnarray}
while the constraints (\ref{Rrho}) read
\begin{eqnarray}\label{RrhoPower}
\rho &=& {\kappa R \over 3(1+w)}{\theta \over y \phi}~~~~,~~~~\phi = {\varepsilon \over 1+\varepsilon} {\theta \over z}, \\ \nonumber \theta &=& 3 (1+w)[ (1+\varepsilon) z +y].
\end{eqnarray}

We use the same process as before to find the fixed points of the dynamical system (\ref{system2}). The results and corresponding physical solutions are depicted in Tables \ref{fix_power1} and \ref{fix_power2}.

\begin{table*}
	\caption{Fixed points and respective solutions for a power law $f(R,\cl)$ function (a).\label{fix_power1}}
	\begin{tabular*}{\textwidth}{@{\extracolsep{\fill}}lc}
		\hline \hline
		Point&$(x,y,z,\phi,\theta)$		\\ \hline 
		
		$\mathcal{A}$	&$\left(1,0,0,0,0 \right)$\\
		
		$\mathcal{B}$	&$\left(x,0,1-x,0,0\right)$\\
		
		$\mathcal{C}$	&$\left({3 \over 2 \varepsilon +1}-1,{3 \over 2 \varepsilon +1}-{1 \over \varepsilon }+2,{1 \over \varepsilon }-{6 \over 2 \varepsilon +1},0,0\right)$\\
		
		$\mathcal{D}$	&$\left(-{3 \varepsilon  (w+1) (3 w-1) \over (\varepsilon +1) \left[6 \varepsilon  (w+1)-3 w-1\right]},{(1-3 w)\over2},{-6 \varepsilon  (w+1)+3 w+1\over2},{3 \varepsilon  (w+1) \left[\varepsilon  \left[6 \varepsilon  (w+1)+3 w+5\right]-2\right] \over (\varepsilon +1) \left[6 \varepsilon  (w+1)-3 w-1\right]},-{3 (w+1) \left[\varepsilon  \left[6 \varepsilon  (w+1)+3 w+5\right]-2\right]\over2}\right)$\\
		\hline \hline
	\end{tabular*}
\end{table*}

\begin{table}
	\caption{Fixed points and respective solutions for a power law $f(R,\cl)$ function (b).\label{fix_power2}}
	\begin{tabular}{lccc}
		\hline \hline
		Point&$a(t)$	&$\rho(t)$	&$q$	\\ \hline 
		
		$\mathcal{A},\mathcal{B}$	&$\left({t \over t_0}\right)^{1 \over 2}$ 	&$\rho_0 \left({t \over t_0}\right)^{-{3 \over 2}(1+w)}$	&$1$ \\
		
		$\mathcal{C}$	&$\left({t \over t_0}\right)^{{\varepsilon  (1+2 \varepsilon) \over 1-\varepsilon}}$ 	&$\rho_0\left({t \over t_0}\right)^{-{3\varepsilon  (1+ 2\varepsilon)(1+w) \over 1-\varepsilon}}$	&$-1+{1 \over \varepsilon } -{3 \over 1+2\varepsilon}$\\
		
		$\mathcal{D}$	&$\left({t \over t_0}\right)^{{2 \over 3(1+w)}}$ 	&$\rho_0 \left({t \over t_0}\right)^{-2}$	&${{1+3w}\over{2}}$	\\
		\hline \hline
	\end{tabular}
\end{table}

\subsection{Point $\ca$}\label{subsec:powerlawA}

\begin{figure}
	\centering
	\includegraphics[width=\columnwidth]{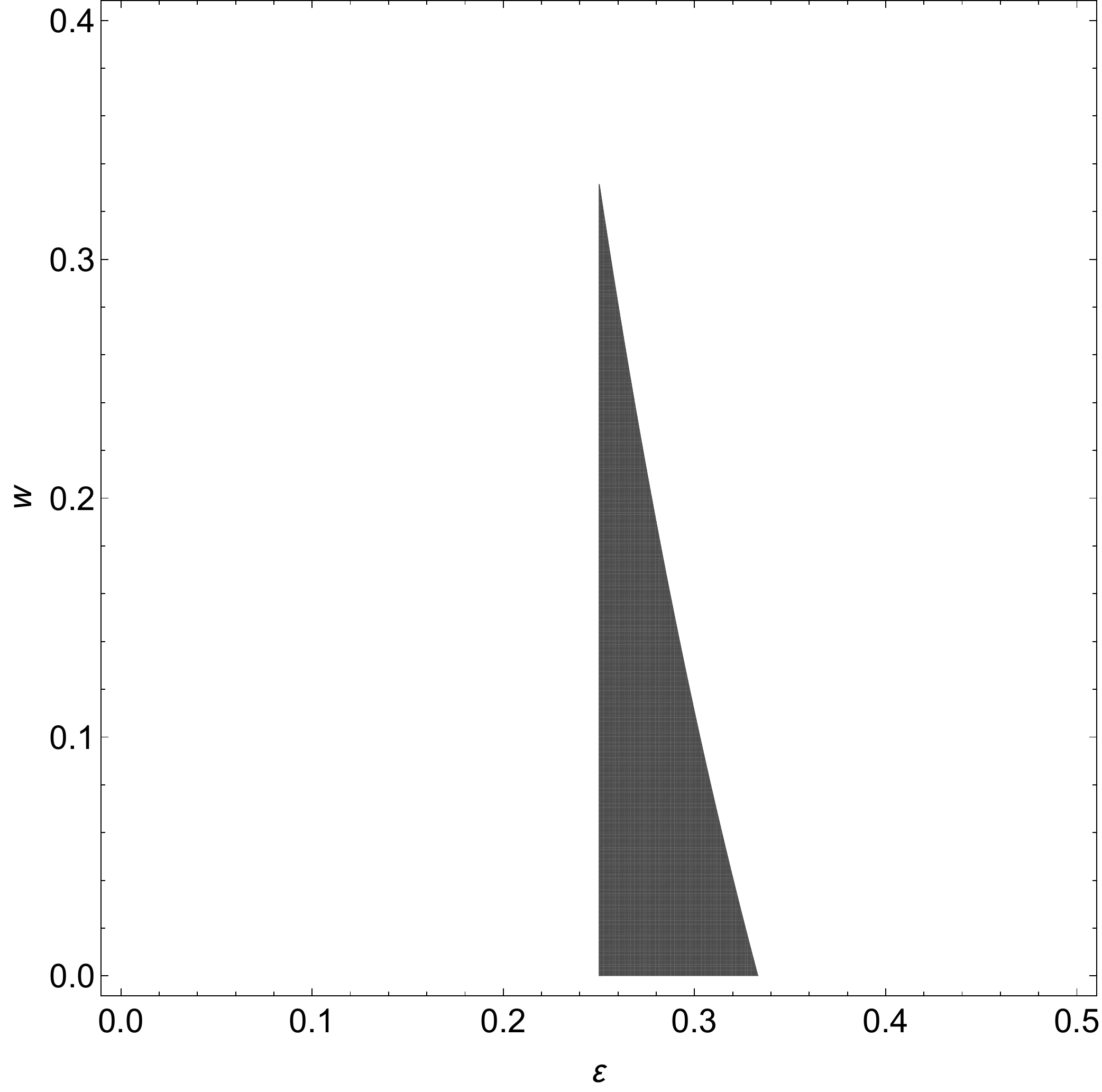}
	\caption{The dark grey region corresponds to the unstable region of point $\ca$. There is no stable region and the remaining phase space corresponds to a saddle point.\label{fig:powerA}}
\end{figure}

Similarly to point $\ca$ in the exponential case, this point is reached asymptotically, since $x=1$ implies that either $f^R=\kappa(1+\varepsilon)(\kappa M^2)^{-\varepsilon}(R-\rho)^\varepsilon=0$ or $H=0$. The scale factor has the same solution as before, so $R$ vanishes and the energy density is inversely proportional to time, and both conditions imply $t\rightarrow\infty$. The stability of the point is shown in Fig. \ref{fig:powerA}.
Note that this fixed point is never an attractor, so it is not a viable candidate for eternal dark energy. It might have some bearing on inflation though, since it does not present a large distinction from GR.

\subsection{Point $\cb$}\label{subsec:powerlawB}

\begin{figure}
	\centering
	\includegraphics[width=\columnwidth]{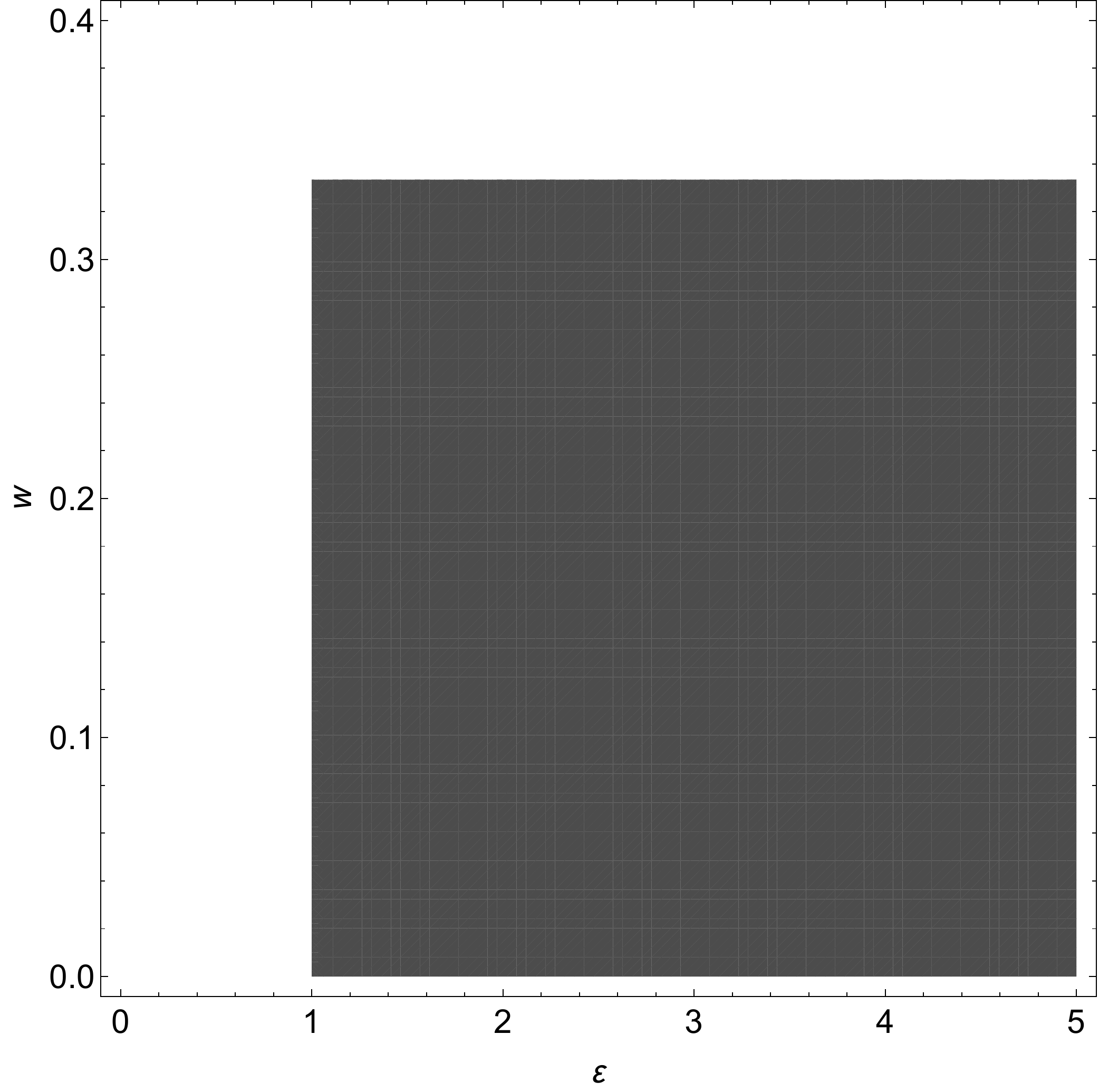}
	\caption{The dark grey region corresponds to the unstable region of point $\cb$. There is no stable region and the remaining phase space corresponds to a saddle point.\label{fig:powerB}}
\end{figure}

Point $\cb$ only occurs if the exponent $\varepsilon$ is related to the EOS parameter $w$ by $\varepsilon=1/3(1+w)$; the equation for the $x$ variable decouples from the rest of the system, so that we have an infinite number of fixed points for all values of $x$, whose stability is shown in Fig. \ref{fig:powerB}. Once again, the point is not an attractor for any values of $x$ and $w$, so it is not a viable candidate for dark energy.
\subsection{Point $\cc$}\label{subsec:powerlawC}

\begin{figure}
	\centering
	\includegraphics[width=\columnwidth]{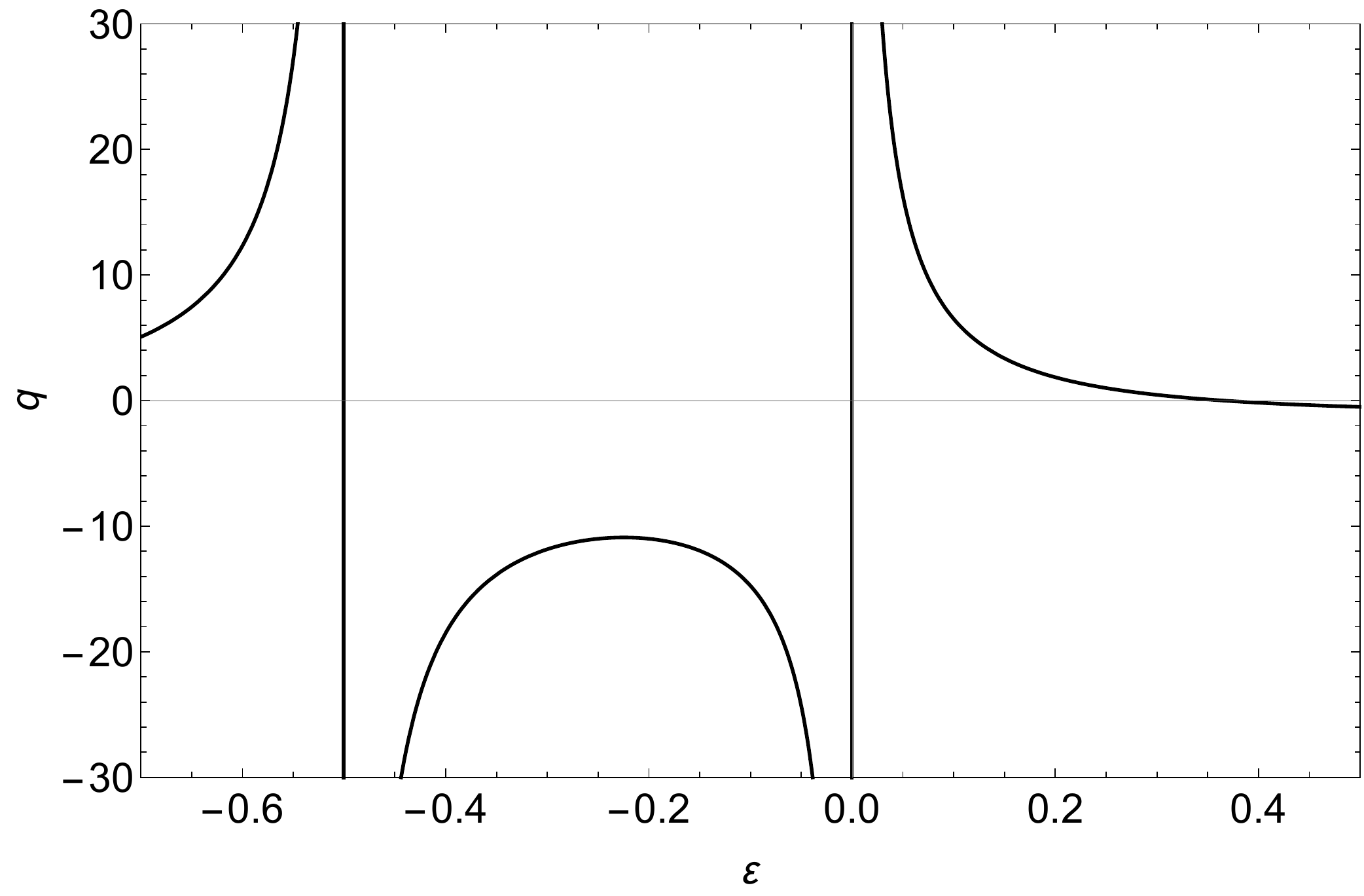}
	\caption{Deceleration parameter for point $\cc$ as a function of the exponent $\varepsilon$.\label{fig:qC}}
\end{figure}

\begin{figure}
	\centering
	\includegraphics[width=\columnwidth]{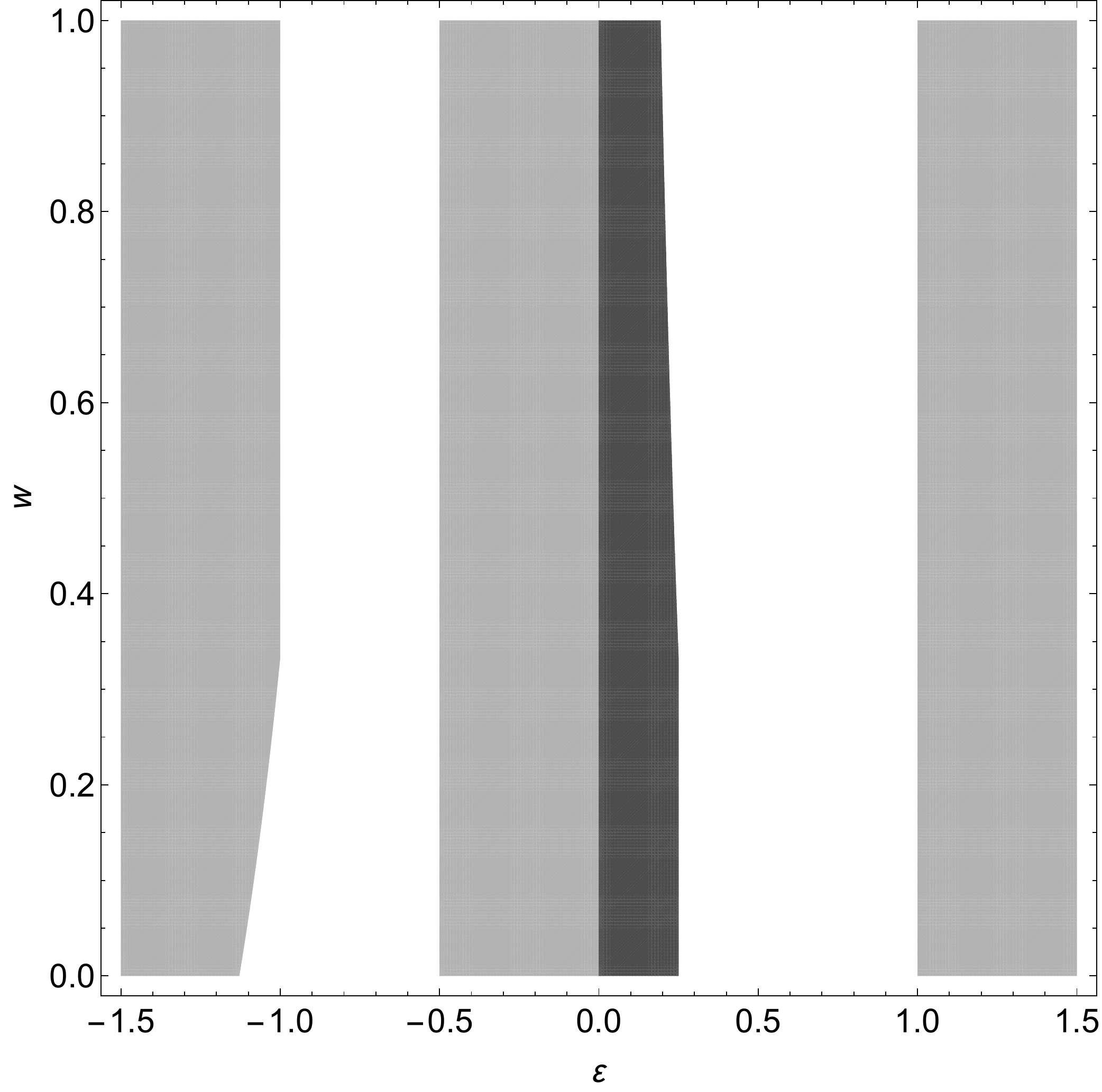}
	\caption{Stability regions of point $\cc$. The dark grey area corresponds to an unstable region, while the light grey corresponds to a stable one. The remaining space corresponds to a saddle point.\label{fig:powerC}}
\end{figure}

This fixed point has the peculiarity of having a deceleration parameter given by
\begin{equation}\label{qC}
q=-1+{1\over \varepsilon } -{3\over 1+2\varepsilon},
\end{equation}
which can be seen more clearly in Fig. \ref{fig:qC}. The stability of the point is shown for a range of $w$ and $\varepsilon$ values in Fig. \ref{fig:powerC}.

Since we are looking for solutions similar to GR, we can exclude the stability regions with $|\varepsilon|>1$ as viable candidates for our theory. The unstable region corresponds to an increasingly higher deceleration parameter as one approaches GR, and as such does not correspond to any known epoch of the Universe.

The stable region with $-1/2< |\varepsilon|<0$, on the other hand, provides a viable alternative to a dark energy, since it is capable of having a negative deceleration parameter for any value of the EOS parameter $w$, and can be arbitrarily close to GR. The only problem that occurs is that the value of $q$ is much lower than one would expect for a dark energy filled Universe, especially as one approaches GR. Even though it leads to a "big rip", this scenario is valid as an asymptotic solution for the current accelerated expansion of the Universe.

\subsection{Point $\cd$}\label{subsec:powerlawD}

\begin{figure}
	\centering
	\includegraphics[width=\columnwidth]{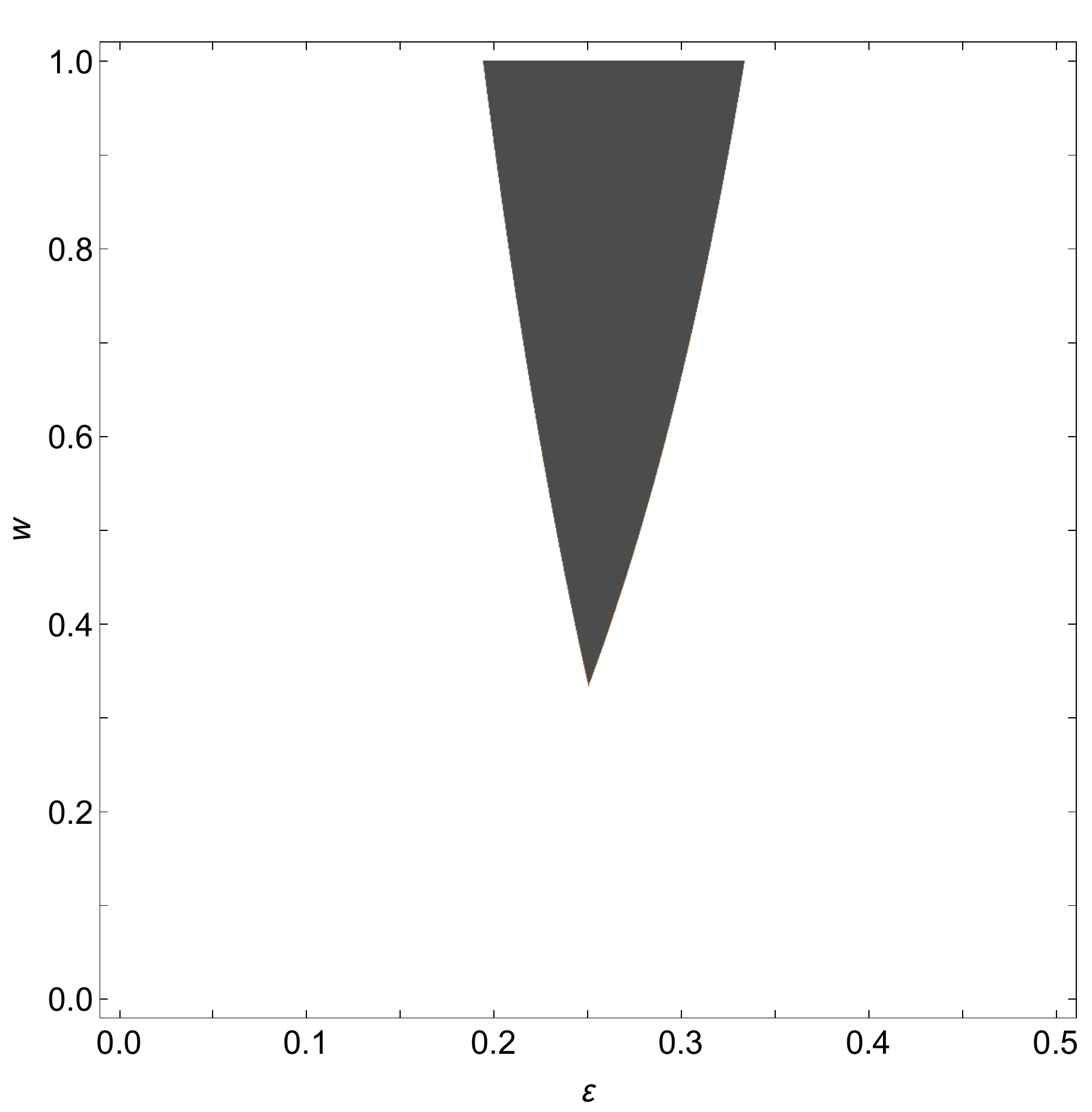}
	\caption{Stability regions of point $\cd$. The dark grey area corresponds to an unstable region and the remaining space corresponds to a saddle point.\label{fig:powerD}}
\end{figure}

This point has a deceleration parameter that is related to the EOS parameter $w$ in the same way as in GR, with $q=(1+3w)/2$. As can be seen in Fig. \ref{fig:powerD}, this point is a saddle point in for all the phase space except for an unstable region with positive $q$, and is therefore an unsuitable candidate for dark energy.

\section{Discussion and outlook}\label{sec:conclusion}

In this work we studied the generic case of $f(R,\cl)$ theories via a dynamical system analysis. As expected, when $f(R,\cl)=\kappa (R-2\Lambda)+\cl$ we obtain the same results as in GR. In the case where the Lagrangian density appears linearly, {\it i.e.} $f(R,\cl)=f_1(R)+f_2(R)\cl$, we obtain the same dynamical system and constraints previously reported in Ref. \cite{rafael2014}.

Forcing a de Sitter solution on the obtained dynamical system shows that one can be obtained for vanishing energy density, as long as the form for the function $f(R,\cl)$ is well behaved, {\it i.e.} does not lead to divergences in the parameters defined in Eq. (\ref{parameters}). Furthermore, we have shown that, as previously explored in NMC theories, de Sitter solutions are possible even for non-vanishing energy density, with the coupling between curvature and matter driving the accelerated expansion of the Universe.

If $f(R,\cl)$ is exponential, we find a saddle point with a radiation-like scale factor solution valid only at $t\rightarrow\infty$, and an unstable fixed point with a de Sitter solution. While the latter is similar to a dark energy filled Universe in GR, the nonminimal coupling of dark energy and curvature lead to this point being unstable.

The power-law form for $f(R,\cl)$ has a wider range of solutions, including a stable fixed point $\cc$ with a negative deceleration parameter that does not require dark energy, making it a possible candidate for an alternative model. The remaining points all have unstable regions, but none that can relate to any known epochs of the history of the Universe.

Overall, dynamical system analysis is a very useful method for checking the compatibility of different gravitation theories against the known history of the Universe. Moreover, the resulting dynamical system is variable-dependent, and as such the choice in the dimensionless variables can affect the range of solutions that can be obtained.

It is also important to remark that the stability of any given solution is only local, and does not imply that there exists a trajectory connecting any two fixed points, as noted in Ref. \cite{Carloni:2007br}. Ideally, one would have a stable fixed point with negative $q$ acting a global attractor, so that all matter-dominated phases would lead to an accelerated expanding Universe without resorting to fine-tuning.

Future work on the cosmological viability of $f(R,\cl)$ theories should include the comparison of the possible candidates for inflation or dark energy here identified with existing cosmographic data, as well as ascertaining under which conditions is this model capable of generating the density perturbations necessary for the creation of large scale structures in the early Universe (see Refs. \cite{Bertolami:2010cw,NMCS} for similar studies in the case of NMC models).

\section*{Acknowledgements}\label{sec:acknowl}

J.P. acknowledges O. Bertolami and F. S. N. Lobo for fruitful discussions. The authors thank the referee for his/her valuable remarks and criticism.

\appendix

\section{Physical quantities}\label{app:physicalq}
Here are listed a few relevant physical quantities in terms of the used dimensionless variables \eqref{variables}. With the adopted metric \eqref{line}, the Ricci scalar reads
\begin{equation}\label{Ricci}
R=6\left(2H^2+\dot{H}\right).
\end{equation}
\noindent One important parameter used in cosmology is the deceleration parameter
\begin{equation}\label{decel}
q \equiv -{\ddot{a}a \over \dot{a}^2}=1-y,
\end{equation}
so that the scalar curvature may be written as
\begin{equation}
R=6H^2(1-q).
\end{equation}
After determining the fixed points of the dynamical system for each particular choice of the function $f(R,\cl)$, we may straightforwardly determine the scale factor for each fixed point. From a direct integration of Eq. \eqref{decel} (for a fixed $y$), one obtains the general solution
\begin{equation}\label{scale}
a(t)=
\begin{cases}
\left({t \over t_0}\right)^{1 \over 2-y}, & y \ne 2 \\
e^{H_0 t}, & y=2
\end{cases}.
\end{equation}
\noindent For the first case, the scale factor evolves as a power of time, while in the second result the Hubble parameter will be constant and this the scale factor will rise exponentially, i.e a De Sitter phase. Note that this solution was obtained resorting (indirectly) to the definition of the Ricci scalar with the used metric.

Other important physical quantity is the energy density: one can determine its evolution for each fixed point from the continuity Eq. \eqref{contin}. The general solution for this is the familiar result
\begin{equation}\label{density}
\rho(t)=\rho_0 a(t)^{-3(1+w)}.
\end{equation}
It is also extremely useful to write both the scalar curvature and the energy density as functions of the variables \eqref{variables}, whose form will depend on the specific model being considered, and thus cannot be fully determined {\it \`a priori}. This can be achieved from the relations
\begin{align}\label{Rrho}
y f &= (x+y+\phi-1)f^R R , \nonumber \\
\theta f&= 3(1+w)(x+y+\phi-1) f^L\rho , \nonumber \\
\theta f^R R &=3(1+w)y f^L \rho, \\
y\phi f&=3(1+w)(1-x-y-\phi)f^{RL}R\rho , \nonumber \\
\phi f^R &=- 3(1+w)f^{RL}\rho, \nonumber \\ \nonumber
y\phi f^L&=-\theta f^{RL}R,
\end{align}
which constitute a system of six equations for the six quantities $(\rho, R, f, f^R, f^L, f^{RL})$. As such, the functions $R=R(x,y,\phi,\theta)$ and $\rho=\rho(x,y,\phi,\theta)$ can only be written explicitly in models in which at least two of the previous equalities are non-trivial, invertible and distinct.

\section{De Sitter Universe}\label{app:desitter}

An interesting if somewhat counter-intuitive use of dynamical analysis is to work backwards and impose a de Sitter Universe, so that the scalar curvature is constant and the energy density vanishes ({\it i.e.} $y=2$), searching for the conditions that the function $f(R,\cl)$ must obey to allow it.

Several quantities depend only on the value of $y$, and can be calculated directly, such as the scale factor $a(t)= e^{H_0 t}$, the density $\rho(t)= \rho_0 e^{-3(1+w)H_0 t}$, and the deceleration parameter $q=-1$. We further assume that $f^R \neq 0$, so that the additional constraint $x = 0 $ is valid, implying that
\begin{equation}\label{Sitterconstraint}\theta=-\phi\left[1 + 3(1+w)\left(\beta_L+1\right) \right].\end{equation}

Substituting the solution for the scale factor into the system, the equation for $y$ simplifies trivially. We are therefore left with the system
\begin{equation}\label{systemSitter}
\begin{cases}
{d\phi \over dN} = \phi\left[\phi - 3(1+w)\left(\beta_L+1\right)\right] \\
{d\theta \over dN} = \theta\left[ \phi-3(1+w)(\gamma_L+1)\right]
\end{cases},
\end{equation}
complemented by the above algebraic constraint, and assuming that $\al_R \neq 0$ and $\beta_R $ does not diverge. 

Since the dependence of $\beta_L$ and $\gamma_L$ on the chosen variables is unknown, we opt not to reduce the above to a single differential equation (with a solution that could not be determined explicitly anyway).

\subsection{Empty Universe solution}

Notice that the assumption of vanishing energy density $\rho = 0 \to \theta = \phi = 0$ was not taken in the above, as the rich phenomenology of $f(R,\cl)$ in principle allows for a relevant contribution from matter which, due to the coupling with curvature, can lead to a de Sitter expansion (as seen in NMC theories \cite{Bertolami:2010cw,Bertolami:2011fz,Bertolami:2013uwl}).

If we nevertheless choose to impose the former condition, then substituting $\theta = \phi = 0$ into the above trivially satisfies both Eq. (\ref{Sitterconstraint}) and the system (\ref{systemSitter}), as long as $\beta_L$ and $\gamma_L$ do not diverge.

As such, we conclude that a ``pure'' de Sitter phase with vanishing energy density is always attainable as long as the function $f(R,\cl)$ does not lead to vanishing parameter $\alpha_R$, $\beta_R$, $\beta_L$ and $\gamma_L$, evaluated at the fixed point $y=2$, $x=\phi=\theta=0$.

\subsection{Non-empty Universe solution}

In the more interesting scenario were a non-vanishing energy density $\rho \neq 0$ nonetheless permits or even drives an exponential phase of accelerated expansion, we may denote the ensuing fixed point(s) as $(x,y,z,\phi,\theta) = (0,2,z^*,\phi^*,\theta^*)$. From Eqs. (\ref{Sitterconstraint},\ref{systemSitter}), we find that the latter must obey
\begin{equation}\label{systemSitternonempty}
\begin{cases}
z^* = -(1+\phi^*), \\
\theta^* = z^* \phi^* \\
\phi^* = 3(1+w)\left[\beta_L(z^*,\phi^*,\theta^*)+1\right] \\
\beta_L(z^*,\phi^*,\theta^*) = \gamma_L(z^*,\phi^*,\theta^*) \neq 0
\end{cases}.
\end{equation}
Notice that, from the definitions (\ref{parameters}), the final condition translates into the condition $f^{LL} f^{RL} = f^L f^{RLL}$ for the function $f(R,\cl)$.

Naturally, since we do not know the explicit dependence of the parameters defined in Eq. (\ref{parameters}) on the dimensionless variables considered, the Jacobian matrix of the dynamical system above cannot be computed, and as such no stability analysis can be performed. Nonetheless, the obtained conditions for the function $f(R,\cl)$ are a relevant result --- as it allows the exclusion of models that do not obey them as suitable dark energy proposals.

\end{document}